\documentclass{KapProc} 
\kluwerbib
\let\lcitebracket(
\let\rcitebracket)
\usepackage[dvips]{graphics,epsfig}
\newcommand{\lsim}{\raise0.3ex\hbox{$<$}\kern-0.75em{\lower0.65ex\hbox{$\sim$}}}
\newcommand{\gsim}{\raise0.3ex\hbox{$>$}\kern-0.75em{\lower0.65ex\hbox{$\sim$}}}

\begin{document}


\articletitle{Semi-analytic galaxy formation: \\
understanding the high redshift universe}
\author{C. M. Baugh$^1$, A. J. Benson$^{2}$, S. Cole$^{1}$, 
C.S. Frenk$^{1}$ and C.G. Lacey$^{3}$.} 
\affil{
$^1$Dept. of Physics, University of Durham, Durham, DH1 3LE, UK.\\
$^2$California Institute of Technology, MC 105-24, Pasadena, CA 91125-2400, USA.\\
$^3$SISSA, via Beirut 2-4, 34014 Trieste, Italy.
}

\begin{abstract}
\noindent
There is now compelling evidence in favour 
of the hierarchical structure formation paradigm. 
Semi-analytic modelling is a powerful tool which 
allows the formation and evolution of galaxies to 
be followed in a hierarchical framework. 
We review some of the latest developments in this 
area before discussing how such models can help us to 
interpret observations of the high redshift Universe.
\end{abstract}


\section*{Hierarchical structure formation}

The hierarchical structure formation paradigm is based upon the 
simple premise that large scale structure in the Universe  
results from the gravitational amplification of small, primordial 
density fluctuations. The origin of the fluctuations is uncertain, 
but one explanation is that they are quantum ripples boosted to 
macroscopic scales by inflation. 

Many clear examples of interacting or merging galaxies, a key 
feature of hierarchical models, were presented during this  
meeting. 
Further convincing evidence for the paradigm can be derived 
by comparing the relative amplitudes of density 
fluctuations in Universe today with those present at some earlier epoch. 
The cosmic microwave background radiation is a snapshot of the 
distribution of photons and baryons just a few hundred thousand 
years after the Big Bang. Fluctuations in the temperature of the 
background radiation can be related to fluctuations in the distribution 
of baryons at the epoch of recombination, $z \sim 1000$. The inferred 
fluctuations are tiny, on the order of one part in a hundred thousand. 
However, if an additional component to the mass density of the Universe 
is included, weakly interacting cold dark matter, these fluctuations 
can subsequently develop into the large scale structure that we measure in the 
Universe today (Peacock et al. 2001). 

An important challenge for theorists is to predict the formation and 
evolution of galaxies in a model universe in which the formation of 
structure in the dark matter proceeds in a hierarchical manner. 
Two powerful simulation techniques have been developed to address this 
issue: direct N-body or grid codes that follow the dynamical 
evolution of dark matter and gas, and semi-analytic codes that use a 
set of simple, physically motivated rules to model the complex physics 
of galaxy formation. 
These techniques have their advantages and disadvantages (e.g. limited 
resolution in case of N-body/grid based codes; the assumption of spherical 
symmetry for cooling gas in the semi-analytics), and so are complementary   
tools with which to attack the problem of 
galaxy formation. 
A preliminary study comparing the cooling of gas and merging of 
``galaxies'' in a Smooth Particle Hydrodynamics simulation with the 
output of a semi-analytic code has shown that there 
is reassuringly good agreement between the results obtained using 
the two techniques (Benson et al. 2001a).

\section*{The Durham semi-analytic code}

The past decade witnessed an explosion in observations of 
galaxies at high redshift, mainly as a result of new facilities such as the 
Hubble Space Telescope and the Keck telescopes in the optical, and 
the opening of other parts of the electromagnetic spectrum, e.g. 
the sub-millimetre, probed by the SCUBA instrument on UKIRT.
In order to interpret these exciting new data, 
semi-analytic galaxy formation codes have been developed that model 
a wide range of physical processes. Below, I will outline the scheme 
developed by the Durham group and collaborators (Benson etal 2000a; 
Cole etal 2000; Granato etal 2000). Similar codes have also been devised  
by other groups (e.g. Avila-Reece \& Firmani 1998; Kauffmann etal 1999; 
Somerville \& Primack 1999). 

The physical processes that play a fundamental role 
in hierarchical galaxy formation can be set out as follows 
(White \& Rees 1978): 

\begin{itemize}
\item[(i)] The formation and merging of dark matter haloes, driven 
by gravitational instability. This process is completely determined 
by the initial power spectrum of density fluctuations and by the values of 
the cosmological parameters $\Omega$, $\Lambda$ and Hubble's constant.
\item[(ii)] The shock heating and virialisation of gas within the gravitational potential 
wells of dark matter haloes.
\item[(iii)] The cooling of gas in haloes. 
\item[(iv)] The formation of stars from cooled gas.
This process is regulated by the injection of energy into the cold gas 
by supernovae and stellar winds.
\item[(v)] The mergers of galaxies after their host dark matter haloes 
have merged. 
\end{itemize}

There are a number of major improvements in the Cole et al. (2000) 
semi-analytic code over earlier versions: a more accurate technique 
is used in the Monte-Carlo generation of dark matter halo merger trees, 
the chemical enrichment of the ISM is followed, disk and bulge scale 
lengths are computed using a prescription based on conservation of 
angular momentum and the obscuration of starlight by dust is computed 
in a self-consistent fashion.

The semi-analytic model requires a number of physical 
parameters to be set. Some of these describe the 
background cosmology and are gradually being pinned down, for example, 
by measurements of supernovae brightnesses at high redshift or 
through the production of high resolution maps of the 
microwave background radiation. 
Other parameters refer to the prescriptions we adopt to model the 
physics of galaxy formation. Their values are set by reference to a subset 
of data on the local galaxy population, as explained by Cole etal.

\begin{figure}
\begin{picture}(300,200)
\put(-5,0)
{\epsfxsize=5.8truecm \epsfysize=6.3truecm 
\epsfbox[80 250 550 600]{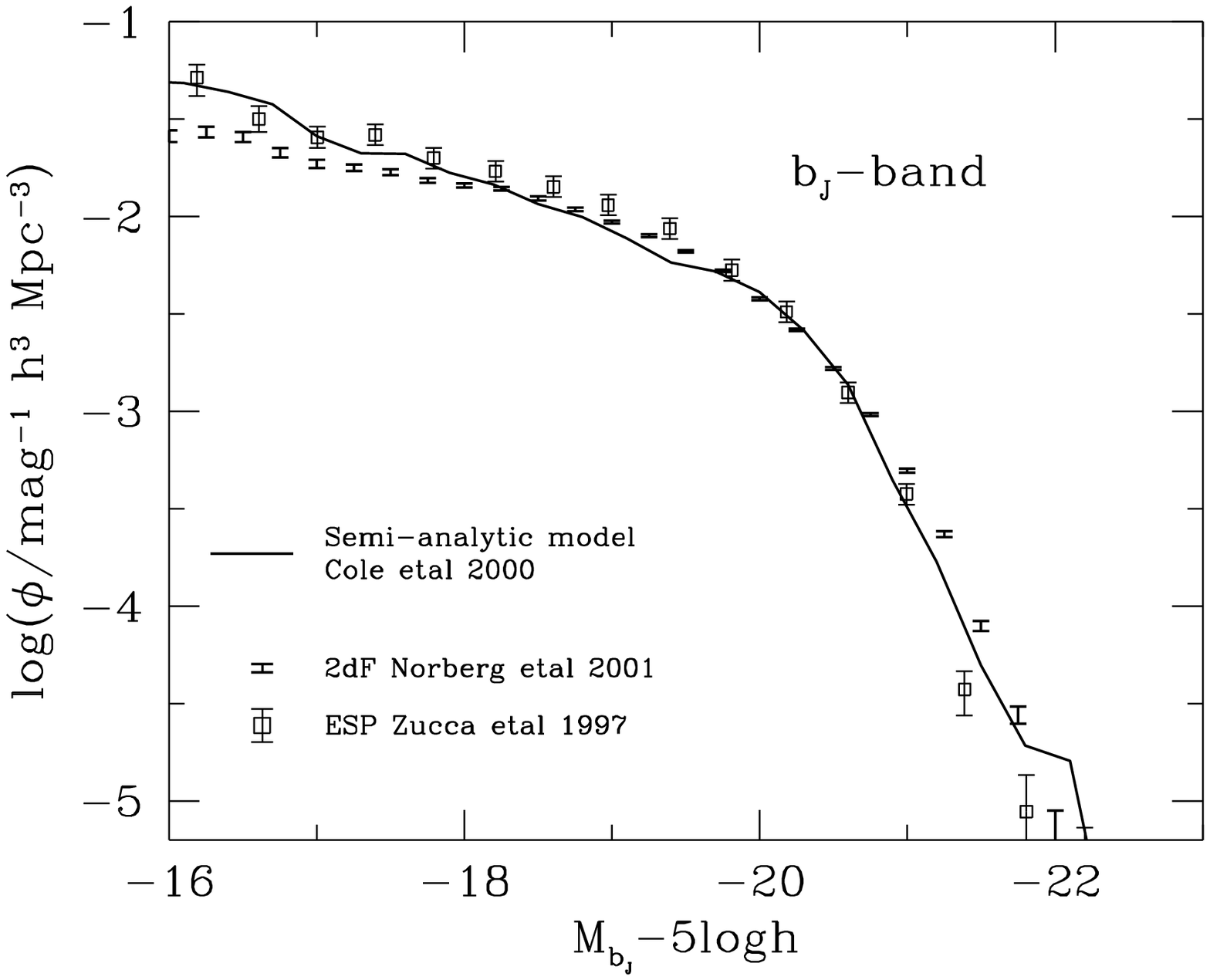}}
\put(170,0)
{\epsfxsize=5.8truecm \epsfysize=6.3truecm 
\epsfbox[80 250 550 600]{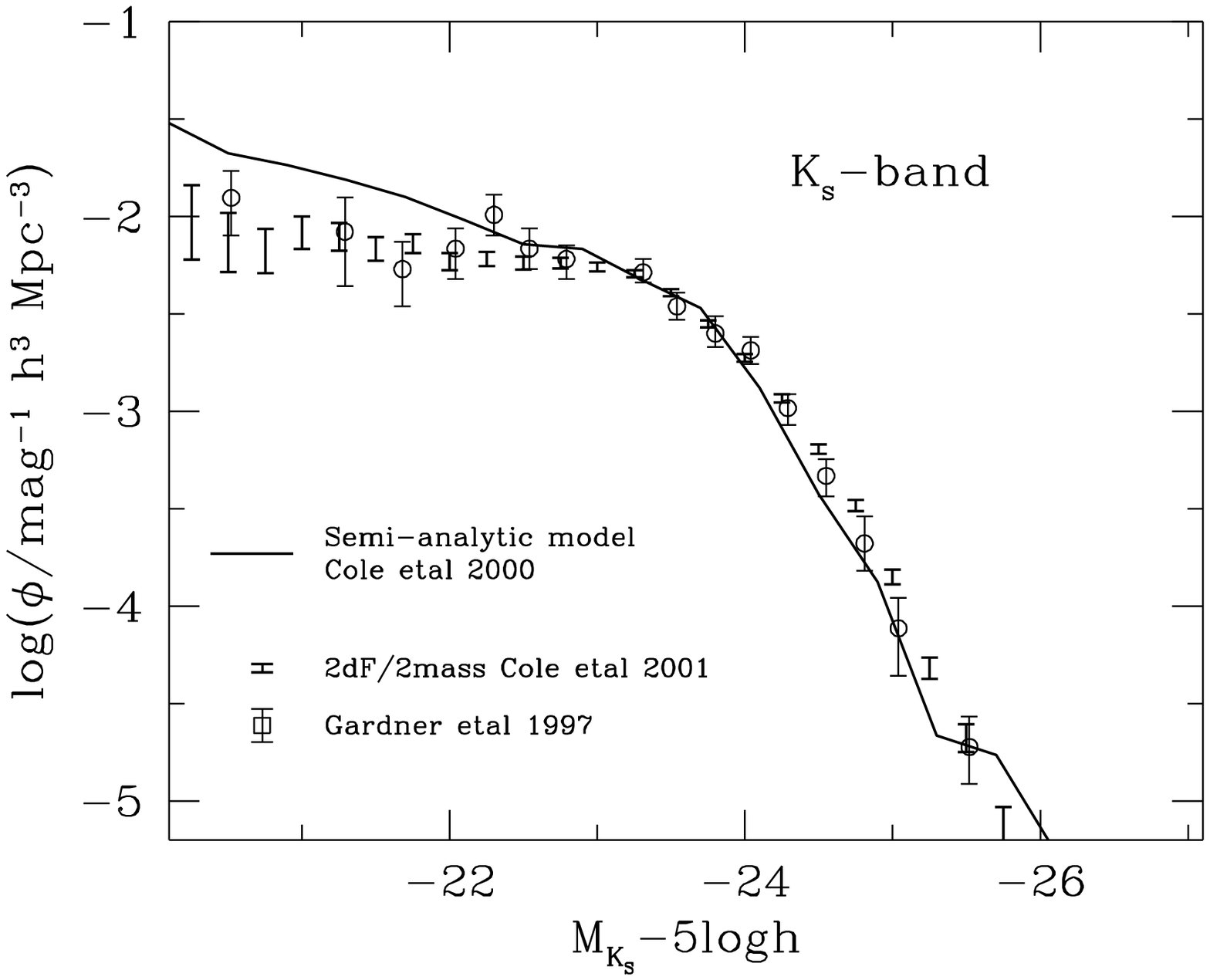}}
\end{picture}
\caption{
The field luminosity function in the $b_{J}$ and $K_{s}$ bands. 
The error bars without points show the luminosity functions 
estimated from the 2dF Galaxy redshift survey (the measurement 
in the right hand panel uses photometry from the near-infrared 
2mass survey).
The points with errorbars show a representative determination 
of the luminosity function from an earlier redshift survey. 
The lines show the luminosity function of the semi-analytic 
model of Cole etal (2000). 
}
\label{fig:lf}
\end{figure}

\begin{figure}
\begin{picture}(100,200)
\put(50,-10)
{\epsfxsize=7.truecm \epsfysize=7.truecm 
\epsfbox[80 270 550 600]{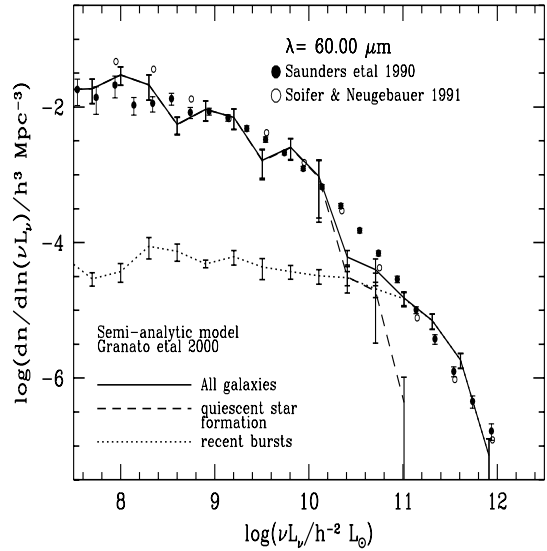}}
\end{picture}
\caption{
The $60\mu m$ luminosity function. The lines show the model 
predictions: the dashed line is the contribution of galaxies 
that are quiescently forming stars, the dotted lines correspond to 
galaxies that recently experienced or are undergoing a burst. The solid 
line is the total luminosity function.
}
\label{fig:lf60}
\end{figure}

The observational constraint to which we attach the most weight is 
the field galaxy luminosity function. Somewhat disappointingly, 
and in spite of much effort, this fundamental characterisation of 
the local galaxy distribution was not well known until this year. 
Fig 5. of Cole et al. (2000) shows that any semblance of a consensus 
between the various determinations of this quantity prior to 2000 
is lost even after moving just one magnitude faintwards of $L_{*}$. 
However, this situation is now changing beyond recognition. 
The 2dF Galaxy Redshift Survey (2dFGRS) and Sloan Digital Sky Survey are 
pinning down the field galaxy luminosity function to a high level of accuracy. 
The degree of improvement that is now possible with the 2dFGRS  
is readily apparent in Fig. \ref{fig:lf}. In this figure, we compare 
measurements obtained from the 2dFGRS with a representative 
determination of the luminosity function made from a redshift survey 
completed in the last millenium. For the first time, random errors 
in the luminosity function estimate are unimportant over a wide range 
of magnitudes.

The solid lines in Fig. \ref{fig:lf} show the luminosity function 
of the Cole etal model. 
The faint end is influenced by the strength 
of feedback in low mass haloes. The break at high luminosities 
is due to long cooling times in more massive dark matter haloes, 
which have higher virial temperatures and form more recently 
in hierarchical models. Assuming a higher galaxy merger rate would depress  
the luminosity function at the faint end and weaken the break at the 
bright end.
From a naive point of view, the model in Fig. \ref{fig:lf} 
would be incorrectly dismissed as  
an abject failure due to an unacceptably large $\chi^2$ value 
with reference to the 2dFGRS estimate of the luminosity function. 
However, it is important to appreciate that the parameters in the 
semi-analytic model are {\it physical} parameters. 
As such, they have a completely different meaning to the 
parameters that specify a Schechter function fit to these data, 
which is merely a convenient mathematical shorthand to describe  
the data points.
We are not at liberty to chose any {\it ad hoc} combination of 
the parameters in the semi-analytic model.
For example, changing the strength of feedback in order to reduce 
the slope of the faint end of the luminosity function also has an 
impact on the shape of the Tully-Fisher relation and upon the 
size of galactic disks.

In collaboration with Alessandro Bressan, Gian-Luigi Granato and 
Laura Silva, we have combined the 
semi-analytic model of Cole etal with the spectro-photometric code of 
Silva et al. (1998), which treats the reprocessing of radiation by dust. 
The range of wavelengths spanned by the spectral energy distribution 
of model galaxies now extends from the extreme ultra-violet through the 
optical to the far-infrared, sub-millimetre and on to the radio 
(Granato et al. 2000).
One highlight of this work is the reproduction of the observed smooth 
attenuation law for starbursts, starting from a dust mixture that reproduces 
the Milky Way extinction law which has a strong bump at $2000\AA$; this 
implies that the observed attenuation is strongly dependent on the 
geometry of stars and dust. 
In Fig. \ref{fig:lf60}, we show the model predictions for 
the $60 \mu m$ luminosity function. Above $\nu L_{\nu} \sim 10^{11} 
h^{-2}L_{\odot}$, the model luminosity function is dominated 
by galaxies undergoing bursts driven by mergers. 
This agrees with observations of ultra-luminous IRAS galaxies, which 
are all identified as being at some stage of the interaction/merger 
process (see Sanders' contribution).

\subsection*{Galaxy clustering at z=0}

\begin{figure}
\begin{picture}(300,160)
\put(-5,0)
{\epsfxsize=6.0truecm \epsfysize=6.truecm 
\epsfbox[80 250 550 580]{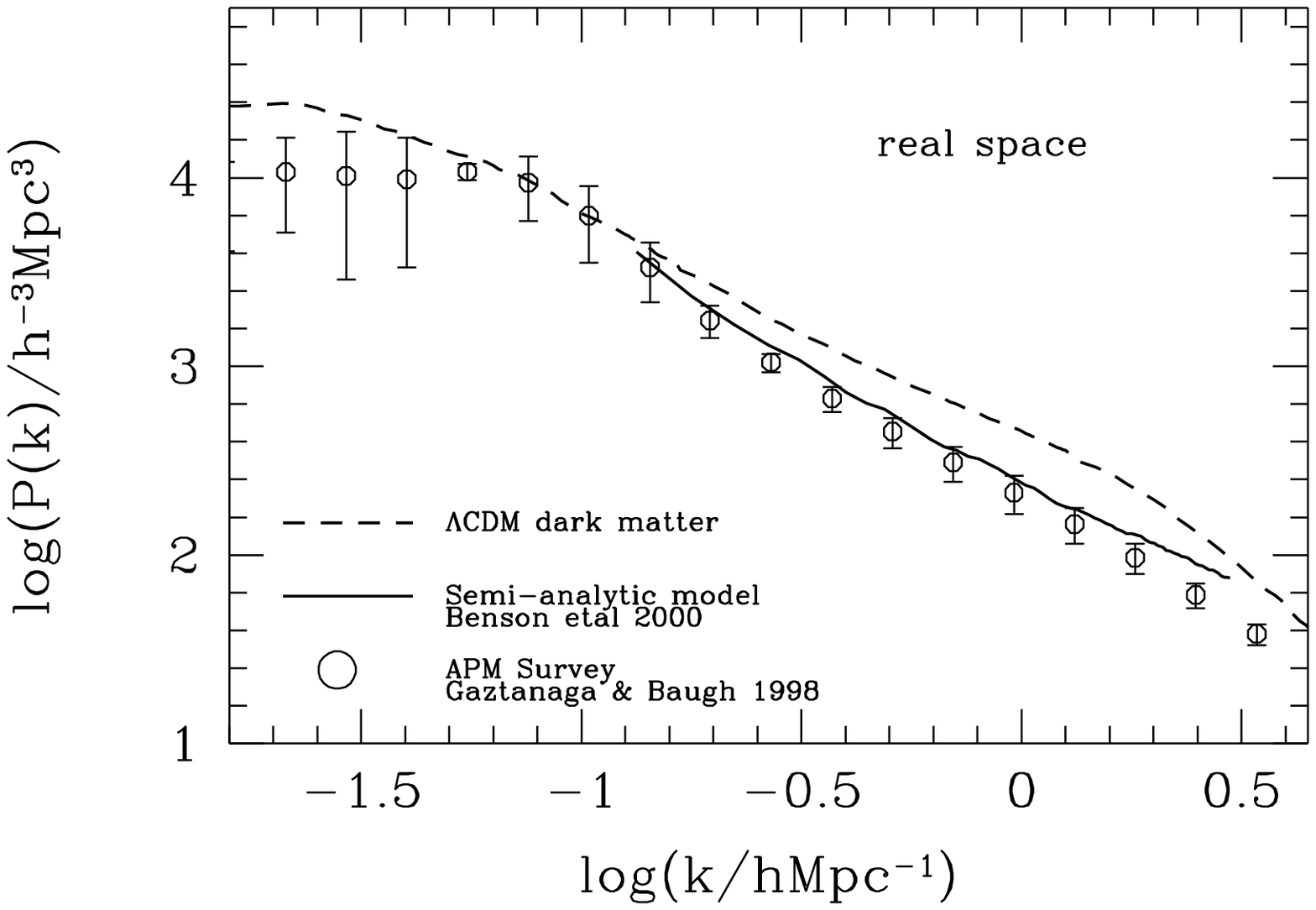}}
\put(170,0)
{\epsfxsize=6.0truecm \epsfysize=6.truecm 
\epsfbox[80 250 550 580]{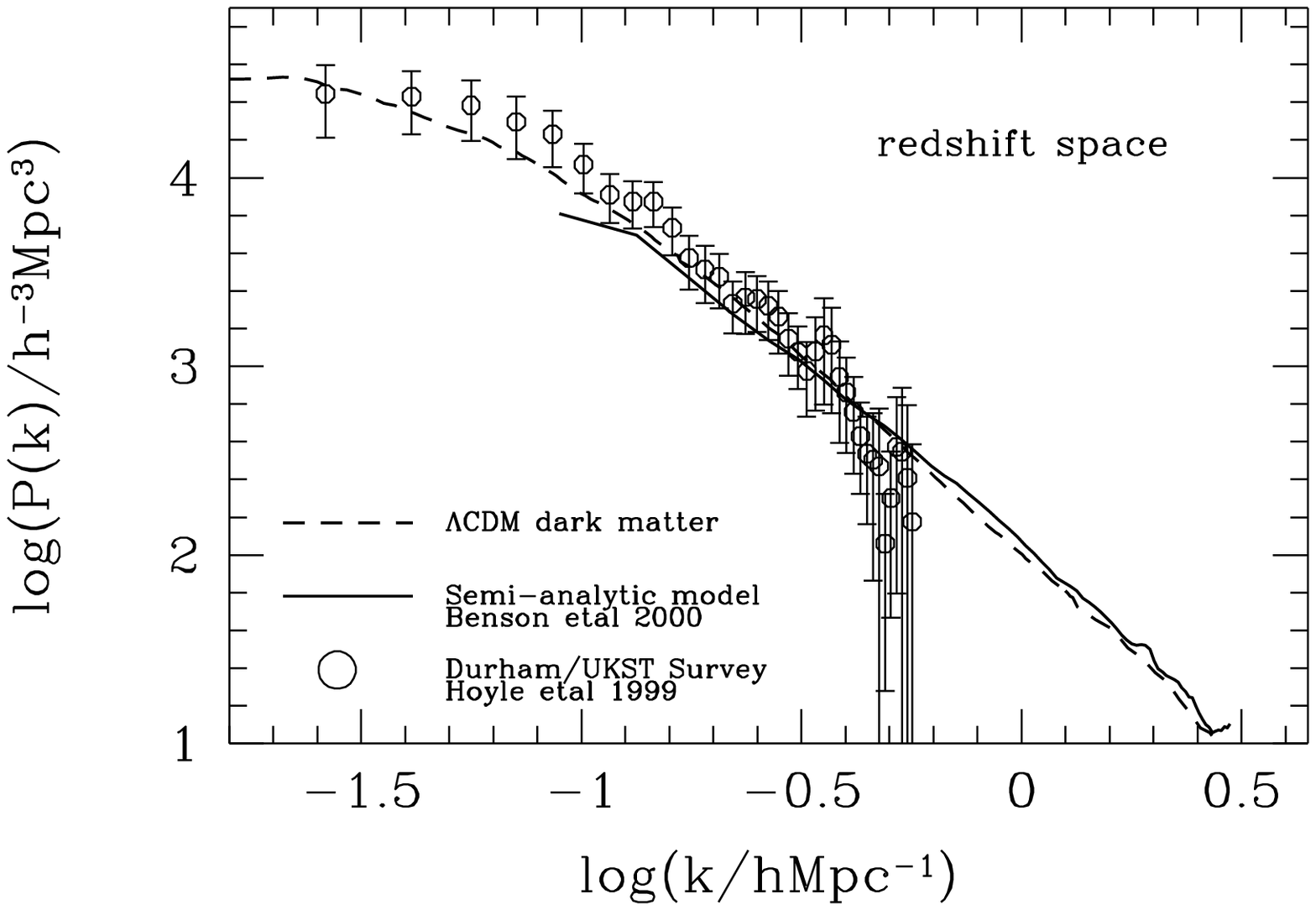}}
\end{picture}
\caption{
The power spectrum of galaxies at $z=0$ in real space (left-hand panel) 
and redshift space (i.e. including the effects of peculiar motions - 
right-hand panel). The solid lines show the predictions for galaxies 
in the semi-analytic models, the dashed lines show the power spectrum 
of the underlying dark matter, which is a CDM universe with 
$\Omega_{0}=0.3$ and $\Lambda_{0}=0.7$. 
The points with errorbars show observational determinations of the 
power spectrum.
}
\label{fig:pk}
\end{figure}

One of the key science goals of the 2dF and SDSS redshift surveys is to 
produce definitive measurements of galaxy clustering over a wide 
range of scales for samples selected by various galaxy properties. 
In order to interpret the information encoded in the measured clustering, 
it is necessary to understand how galaxies illuminate the underlying 
distribution of dark matter.
Progress has been made towards this end by marrying the 
semi-analytic galaxy formation technique with high resolution N-body 
simulations of representative volumes of the universe 
(Kauffmann et al. 1999; Benson et al. 2000a,b, 2001b).

In the approach of Benson et al., the masses and positions of dark matter 
haloes are extracted from an N-body simulation using a standard group 
finding algorithm. The semi-analytic machinery is then employed to 
populate the dark haloes with galaxies. The central galaxy is placed 
at the centre of mass of the dark matter halo and satellite galaxies 
are placed on random dark matter particles within the halo, resulting in a 
map of the spatial distribution of galaxies within the simulation volume.
Fig. \ref{fig:pk} compares the power spectrum of bright, optically 
selected galaxies predicted by the semi-analytic model, with   
observational determinations and with the power spectrum of the dark matter. 
The left hand panel shows power spectra in real space. For  
$k\gsim 0.1 h {\rm Mpc}^{-1}$, the measured galaxy power spectrum 
has a lower amplitude than that of the dark matter in the popular 
$\Lambda$CDM model; the galaxies are said to be `anti-biased' with 
respect to the mass (Gazta\~{n}aga 1995). The semi-analytic model 
provides an excellent match to the data. This is particularly 
noteworthy as no additional tuning of parameters was carried out 
to make this prediction once certain properties of the local 
galaxy population, such as the field galaxy luminosity function, 
had been reproduced (see Cole et al. 2000 for a full explanation of 
how the model parameters are set). Furthermore, this level of agreement 
is not found for the galaxy clustering predicted in CDM models with $\Omega=1$.
The most important factor in shaping the predicted galaxy clustering 
amplitude is the way in which the efficiency of galaxy formation depends 
upon dark matter halo mass.
This is illustrated by the variation in the mass to light ratio with 
halo mass shown by Fig 8 of Benson et al (2000a): for low mass haloes, galaxy 
formation is suppressed by feedback, whilst for the most massive haloes, 
gas cooling times are sufficiently long to suppress cooling.

The power of the approach of combining semi-analytic models with N-body 
simulations is demonstrated on comparing the left hand panel (real space) of 
Fig. \ref{fig:pk} with the right hand panel, which shows power spectra 
in redshift space. Again, the same model gives a very good match to the 
observed power spectrum when the effects of peculiar motions are included 
to infer galaxy positions. However, the impression that one would gain 
about the bias between dark matter and galaxy fluctuations is qualitatively 
different; in redshift space galaxies appear to be unbiased tracers of the dark 
matter. The apparent contradiction between the implications for bias given 
by the panels of Fig. \ref{fig:pk} can be resolved by turning back once more 
to the models. The pairwise velocity dispersion of model galaxies 
is lower than that of the dark matter, and as a result is in much 
better agreement with the observational determination of pairwise motions. 
Again, this difference is driven by a reduction in the efficiency of 
galaxy formation with increasing dark matter halo mass (Benson et al. 2000b).

\subsection*{The evolution of the galaxy distribution}

\begin{figure}
{\epsfxsize=10.0truecm \epsfysize=6.truecm 
\epsfbox[-100 250 575 580]{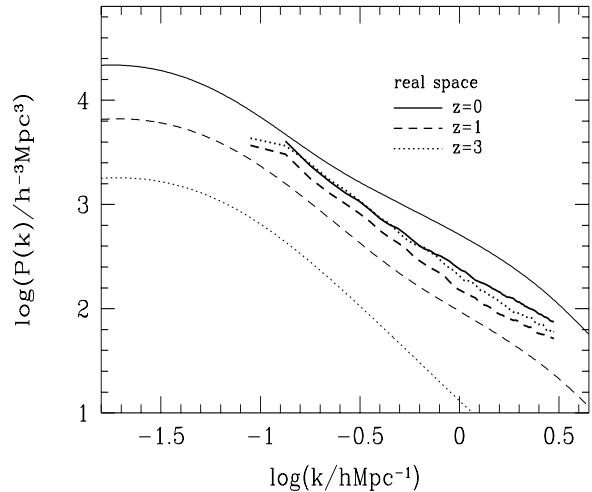}}
\caption{
The evolution of galaxy clustering. The thick lines show 
the power spectrum predicted by the semi-analytic model 
for galaxies brighter than $L_{*}$ (defined as 
$M_{b_{J}}-5\log h=-19.5$ in the rest frame $b_{J}$-band)
at $z=0,1$ and $3$.
The light lines show the power spectrum of the dark matter 
at the same epochs, reading from top to bottom. 
}
\label{fig:pkz}
\end{figure}

Once the parameters of the semi-analytic model have been set by 
comparing the model output with a subset of data for the local 
galaxy population, firm predictions can be made regarding the 
evolution of the galaxy distribution (Benson et al. 2001b).

The properties of the distribution of galaxies and the way in which 
these properties evolve with redshift are intimately connected to 
the growth of structure in the dark matter, as illustrated by a 
sequence of high resolution pictures in Benson et al. (2001b) that 
show the evolution of galaxies and of the dark matter.
An example of this is the morphology-density relation, namely the 
correlation of the fraction of early type galaxies with local 
galaxy density. The semi-analytic models reproduce the observed form 
of the morphology density relation at $z=0$. Remarkably, essentially the 
same strength of effect is also predicted at $z=1$. The physical 
explanation for this result lies in the accelerated dynamical evolution 
experienced by galaxies that form in overdensities destined to become rich 
clusters by the present day. 

A generic prediction of hierarchical clustering models is that 
bright galaxies should be strongly clustered at high redshift 
compared to the underlying dark matter (Davis et al. 1985). 
Fig. \ref{fig:pkz} shows the 
evolution of the power spectrum for galaxies and for dark matter in a 
$\Lambda$CDM universe. 
The amplitude of the dark matter power spectrum increases as fluctuations 
grow through gravitational instability. Between $z=3$ and $z=0$, the 
amplitude of the dark matter power spectrum increases by an order of 
magnitude on large scales. The shape of the dark matter power spectrum 
is significantly modified on small scales (high $k$) through nonlinear 
evolution of the density fluctuations -- `cross-talk' between fluctuations 
on different spatial scales. However, the amplitude and shape of the 
galaxy power spectrum show little change over the same redshift interval 
(Pearce et al. 2000; Benson et al. 2001b). The amplitude of the 
galaxy power spectrum drops by around $50\%$ from $z=3$ to $z=1$, 
and by $z=0$ it has been overtaken in amplitude by the mass power 
spectrum (Baugh et al. 1999). 
The clustering predictions can be readily explained. At $z=3$,  
bright galaxies are only found in the most massive haloes  
in place at this time. Such haloes are much more strongly clustered  
than the underlying dark matter, hence the large difference in amplitude 
or bias between the galaxy and dark matter spectra 
at $z=3$. The environment of bright galaxies becomes less exceptional 
as $z=0$ is approached.

\subsection*{The formation and evolution of QSOs}

The similarity in the general evolution of the global star formation 
rate per unit volume and of the space density of luminous quasars 
suggests a connection between the physical processes that drive 
the formation and evolution of galaxies and those 
that power quasars (see Dunlop's contribution). 
Spurred on by mounting dynamical evidence for the presence of 
massive black holes in galactic bulges (e.g. Magorrian 
et al. 1998), Guinevere Kauffmann and Martin Haehnelt 
have produced the first treatment to follow the properties of QSOs 
within a fully fledged semi-analytic model for galaxy formation 
(Kauffmann \& Haehnelt 2000; Haehnelt \& Kauffmann 2000).

The model of Kauffmann \& Haehnelt assumes that black 
holes form during major mergers of galaxies, and 
that during the merger event, some fraction of the cold gas 
present is accreted onto the black hole to fuel a quasar. 
The qualitative properties of the observed quasar population 
are reproduced well by the model, including the rapid evolution 
in the space density of luminous quasars. 
There are three key features of the model responsible for 
the evolution in quasar space density between $z\sim 2$ and 
$z=0$: (i) a decrease in the merger rate of objects in a fixed 
mass range over this interval, (ii) a reduction in the supply of 
cold gas from mergers, and (iii) an increase in the time-scale for 
gas accretion onto the black hole. 
The mass of cold gas available in mergers is reduced at low redshift 
because the star formation timescale in the model is effectively 
independent of redshift; at lower redshifts, gaseous disks 
have been in place for longer and a larger fraction of the gas has been 
consumed in quiescent star formation and so less gas is present in 
low redshift mergers (see Fig 6 of Baugh, Cole \& Frenk 
1996). If the star formation timescale is allowed to depend upon the 
dynamical time, gas is consumed more rapidly in the disk and less gas 
is present in mergers at all redshifts. 

The Kauffmann \& Haehnelt model predicts strong evolution in the 
properties of QSO hosts with redshift, suggesting that quasars 
of a given luminosity should be found in fainter hosts at high  
redshift. This issue is just beginning to be addressed observationally 
(see, for example, the contributions of Ridgway and Kukula). 
At present, it is hard to reach any firm 
conclusions, though there is apparently little evidence for a strong 
trend in host luminosity with redshift. 

The 2dF QSO redshift survey has recently reported measurements of the 
clustering in a sample of QSOs that is an order of magnitude larger than 
any previous sample (Hoyle et al. 2001). It should be relatively 
straight forward to obtain predictions for the clustering of quasars 
from the semi-analytic models to compare with these new data.

\subsection*{Acknowledgments}

CMB would like to thank the organisers of this enjoyable meeting 
for their hospitality and for providing financial support.   
We acknowledge the contribution of our GRASIL collaborators, 
Alessandro Bressan, Gian-Luigi Granato and Laura Silva 
to the work presented in this review. We thank Peder Norberg 
and the 2dF Galaxy Redshift Survey team for communicating preliminary 
luminosity function results.


\end{document}